\documentclass[traditabstract]{aa} 
\usepackage{graphicx}
\usepackage{txfonts}
\usepackage{natbib}
\usepackage{wasysym} 
\bibpunct{(}{)}{;}{a}{}{,}
\newcommand{\teff}{\mbox{$T_{\rm eff}$}}
\newcommand{\teq}{\mbox{$T_{\rm eq}$}}
\newcommand{\logg}{\mbox{$\log g$}}
\newcommand{\vsini}{\mbox{$v \sin i^{\star}$}}
\newcommand{\mictrb}{\mbox{$\xi_{\rm t}$}}
\newcommand{\mactrb}{\mbox{$v_{\rm mac}$}}

\newcommand{\kms}{\mbox{km\,s$^{-1}$}}
\newcommand{\halpha}{\mbox{$H_\alpha$}}
\newcommand{\mura}{\mbox{$\mu_\alpha$}}
\newcommand{\mudec}{\mbox{$\mu_\delta$}}

\newcommand{\rhostar}{\ensuremath{\rho_\star}}
\newcommand{\rhosun}{\ensuremath{\rho_\odot}}
\newcommand{\rhoj}{\ensuremath{\rho_{\rm J}}}
\newcommand{\rhopl}{\ensuremath{\rho_{\rm pl}}}
\newcommand{\rj}{R\ensuremath{_{\rm J}}}
\newcommand{\mj}{M\ensuremath{_{\rm J}}}
\newcommand{\rsun}{R\ensuremath{_\odot}}
\newcommand{\msun}{M\ensuremath{_\odot}}
\newcommand{\rpl}{\ensuremath{R_{\rm pl}}}
\newcommand{\mpl}{\ensuremath{M_{\rm pl}}}
\newcommand{\rstar}{\ensuremath{R_\star}}
\newcommand{\mstar}{\ensuremath{M_\star}}

\def\secos{$\sqrt{e} \cos \omega$}
\def\sesin{$\sqrt{e} \sin \omega$}
\def\feh{[Fe/H]}

\begin{document}
   \title{WASP-39b: a highly inflated Saturn-mass planet orbiting a late G-type star}

   \author{F. Faedi \inst{1} \and S. C. C. Barros \inst{1} \and D.
     R. Anderson \inst{2} \and D. J. A. Brown \inst{3} \and A. Collier
     Cameron \inst{3} \and D. Pollacco \inst{1} \and I. Boisse
     \inst{4,5} \and G. H\'ebrard \inst{4,6} \and M. Lendl \inst{7}
     \and T. A. Lister \inst{8} \and B. Smalley \inst{2} \and R. A.
     Street \inst{8} \and A. H. M. J. Triaud \inst{7} \and J. Bento
     \inst{9} \and O. W. Butters \inst{10} \and B. Enoch \inst{3}
     \and F. Bouchy \inst{4,6} \and C. A. Haswell \inst{11}
     \and C. Hellier \inst{2} \and F. P. Keenan \inst{1} \and G. R. M.
     Miller \inst{3} \and V. Moulds \inst{1}  \and C. Moutou
     \inst{12}\and A. J. Norton \inst{11} \and D. Queloz \inst{7} \and
     A. Santerne \inst{4,6} \and E. K. Simpson \inst{1} \and I.
     Skillen \inst{13} \and A. M. S. Smith \inst{2} \and S. Udry
     \inst{7} \and C. A. Watson \inst{1} \and R. G. West \inst{10}
     \and P. J. Wheatley \inst{9}}

\institute{Astrophysics Research Centre, School of Mathematics and Physics, Queen's University Belfast, University Road, Belfast BT7 1NN.\\
  \email{f.faedi@qub.ac.uk}  
\and Astrophysics Group, Keele University,  Staffordshire, ST5 5BG, UK \\ 
\and School of Physics and Astronomy,  University of St Andrews, St Andrews, Fife KY16 9SS, UK \\ 
\and Institut d'Astrophysique de Paris, UMR7095 CNRS, Universit\'e Pierre \& Marie Curie, France \\
\and Centro de Astrof\'isica, Universidade do Porto, Rua das Estrelas, 4150-762 Porto, Portugal \\
\and Observatoire de Haute-Provence, CNRS/OAMP, 04870 St Michel l'Observatoire, France \\
\and Observatoire astronomique de l' Universit\'e de Gen\'eve, 51 ch.\ des Maillettes, 1290 Sauverny, Switzerland \\
\and Las Cumbres Observatory Global Telescope Network, 6740 Cortona Drive Suite 102, CA 93117, USA\\ 
\and Department of Physics, University of Warwick, Coventry CV4 7AL, UK \\
\and Department of Physics and Astronomy, University of Leicester, Leicester, LE1 7RH \\
\and Department of Physics and Astronomy, The Open University, Milton Keynes, MK7 6AA, UK \\
\and Laboratoire d'Astrophysique de Marseille, 38 rue Fr\'ed\'eric Joliot-Curie, 13388 Marseille cedex 13, France \\
\and Isaac Newton Group of Telescopes, Apartado de Correos 321, E-38700 Santa Cruz de Palma, Spain}

   \date{Received; accepted }

  \abstract
  {We present the discovery of WASP-39b, a highly inflated transiting
    Saturn-mass planet orbiting a late G-type dwarf star with a period
    of $4.055259 \pm 0.000008$\,d, Transit Epoch
    T$_{0}=2455342.9688\pm0.0002$\,(HJD), of duration
    $0.1168 \pm 0.0008$\,d. A combined analysis of the WASP photometry,
    high-precision follow-up transit photometry, and radial velocities
    yield a planetary mass of $\mpl=0.28\pm0.03\,\mj$~and a radius of
    $\rpl=1.27\pm0.04\,\rj$, resulting in a mean density of
    $0.14 \pm 0.02\,\rhoj$. The stellar parameters are mass 
    $\mstar = 0.93 \pm 0.03\,\msun$, radius $\rstar = 0.895\pm 0.23\,\rsun$,
    and age $9^{+3}_{-4}$\,Gyr. Only WASP-17b and WASP-31b have lower
    densities than WASP-39b, although they are slightly more massive
    and highly irradiated planets. From our spectral
    analysis, the metallicity of WASP-39 is measured to be
    \feh\,$= - 0.12\pm0.1$\,dex, and we find the planet to have an
    equilibrium temperature of $1116^{+33}_{-32}$\,K\,. Both values
    strengthen the observed empirical correlation between these
    parameters and the planetary radius for the known transiting
    Saturn-mass planets. \thanks{Spectroscopic and photometric data
      are available in electronic form at the CDS via anonymous ftp to
      cdsarc.u-strasbg.fr (130.79.128.5) or via
      http://cdsweb.u-strasbg.fr/cgi-bin/qcat?J/A+A/}}


   \maketitle
%

\section{Introduction}

The importance of transiting extrasolar planets is related to their
geometrical configuration \citep{Sackett99}. Transit geometry severely
constrains the orbital inclination of the planet, allowing accurate
measurements of its mass and radius to be derived. The inferred
planet's density provides information on the system's bulk physical
properties, and thus is a fundamental parameter for constraining
theoretical models of planetary formation, structure and evolution
(e.g. \citealt{Guillot05}; \citealt{Fortney07}; \citealt{Liu08}).

To date, more than 100 transiting planets have been discovered, which
show a huge range of diversity in their physical and dynamical
properties. For example, their mass ranges from $\sim5\,M_{\oplus}$
(Kepler-10b, \citealt{Batalha11}) to about 12\,\mj\ (XO-3b,
\citealt{Johns-Krull08}; and \citealt{Hebrard08}). Some planets have
radii that agree with models of irradiated planets
(\citealt{Burrows07}; \citealt{Fortney07}), while others are found to
be anomalously large (e.g. WASP-12b \citealt{Hebb09} and TrES-4b
\citealt{Southworth10}, \citealt{Torres08}, \citealt{Mandushev07}).
The diversity in exoplanet densities, and hence in their internal
compositions, is particularly noticeable at sub-Jupiter masses. For
example, some exoplanets have very high densities and are thought to
have a rocky/ice core (e.g. HD 149026b,~$\rhopl\simeq 1\,\rhoj$,
\citealt{Sato05}), while systems such as TrES-4b ($\rhopl=
0.17\,\rhoj$, \citealt{Mandushev07}), WASP-17b ($\rhopl =
0.06\,\rhoj$,~\citealt{Anderson10a}; \citealt{Anderson11b}), WASP-31b
($\rhopl = 0.132\,\rhoj$, \citealt{Anderson10b}), and Kepler-7b
($\rhopl = 0.13\,\rhoj$, \citealt{Latham10}) are examples of planets
with puzzlingly low densities that challenge standard evolutionary
theories in reproducing their radii (\citealt{Fortney07},
\citealt{Burrows07}). To assess the inflation status of a system,
generally planetary radii are compared to tabulated values from models
(e.g. \citealt{Fortney07}, \citealt{Burrows07}, \citealt{Baraffe08}).
However, the radius depends on multiple physical properties, such as
the stellar age, the irradiation flux, the planet's mass and age, the
atmospheric composition, the presence of heavy elements in the
envelope or in the core, the atmospheric circulation, and also on any
source generating extra heating in the planetary interior. Different
mechanisms have been proposed to explain the anomalously large radii,
such as tidal heating due to unseen companions pumping up the
eccentricity \citep{Bodenheimer01, Bodenheimer03}, kinetic heating due
to the breaking of atmospheric waves \citep{Guillot02}, enhanced
atmospheric opacity \citep{Burrows07} and semi-convection
\citep{Chabrier07}. While each individual mechanism would presumably
affect all hot Jupiters to some degree, they cannot explain the
entirety of the observed radii (\citealt{Fortney10},
\citealt{Baraffe10}). More complex thermal evolution models are
necessary to fully understand their cooling history. For these
planets, the dominant source of energy is a function of the orbital
separation and the spectral type of the host star, while its
dependency on the planetary mass and age is negligible (with the
exception of very young planets). More recently, \citet{Batygin11}
performed calculations of the thermal evolution of gas giants and
suggest that the extra energy needed to explain the radius inflation
comes from stellar irradiation (see also \citealt{Laughlin11}). The
proposed mechanism (Ohmic heating), based on the interaction of
ionised alkali metals in the planetary atmosphere with the planet's
magnetic field, along with the atmospheric heavy element content,
could provide a universal explanation of the currently measured radius
anomalies.

Although the majority of the known exoplanets are short-period,
Jupiter-mass planets, more recently an increasing number of
Saturn-like planets have been discovered (e.g. \citealt{Enoch10}),
and have encouraged studies of planetary properties and their
statistical analysis, searching for possible correlations between
planetary parameters (e.g. \citealt{Enoch10}; \citealt{Anderson10b};
\citealt{Hartman10}). To date, 27 transiting planets have been
discovered with masses in the range $0.15\,\mj<\mpl<0.6\,\mj$
\footnote{http://exoplanet.eu/}, similar to Saturn ($M_{\rm
  Saturn}=0.229\,\mj$, \citealt{Standish95}). The detection and
characterisation of significantly more bright short-period
transiting systems is one of the keys to improving our understanding
of planetary structure and evolution.

Here we describe the properties of WASP-39b, a new transiting
Saturn-mass planet discovered by the SuperWASP survey. The planet
host star WASP-39 belongs to the constellation of Virgo and thus
resides in an equatorial region of sky, which is monitored by the
SuperWASP-North and WASP-South telescopes simultaneously. WASP-39b
is the third least dense planet ($\rhopl = 0.14\pm0.02\,\rhoj$)
discovered from a ground-based transit survey, and belongs to the
sample of highly inflated gas giant planets. It provides
observational evidence for the mass-radius relation of planetary
systems in a poorly sampled region of the parameter space.

We present follow-up observations of the new system which establish
the planetary nature of the transiting object detected by SuperWASP.
High precision, high signal-to-noise light-curves have been obtained
using both the Faulkes Telescope North (FTN), and the Euler
telescope, and radial velocity measurements using the SOPHIE
(1.93--m OHP) and CORALIE (Swiss 1.2--m) spectrographs.

The paper is structured as follows: in $\S$\,2 we describe the
observations, including the WASP discovery data and the photometric
and spectroscopic follow-up. In $\S$\,3 we present our results for
the derived systems parameters, and the stellar and planetary
properties. Finally, we discuss the implication of the discovery of
WASP-39b in $\S$\,4.


\section{Observations}

1SWASP J142918.42-032640.1 (2MASS 14291840-0326403), hereafter
WASP-39, has been identified in several northern sky catalogues which
provide broad-band optical \citep{Nomadcat} and infra-red 2MASS
magnitudes \citep{2MASS} as well as proper motion information.
Coordinates, broad-band magnitudes, and proper motion of the star are
from the NOMAD catalogue and are given in Table~\ref{table1}.

\subsection{SuperWASP observations}

The WASP North and South telescopes are located in La Palma (ING -
Canaries Islands) and Sutherland (SAAO - South Africa), respectively.
Each telescope consists of 8 Canon 200mm f/1.8 focal lenses coupled to
e2v $2048 \times 2048$ pixel CCDs, yielding a field-of-view of
$7.8 \times 7.8$ square degrees with a pixel scale of 13.7\arcsec ~\citep{Pollacco06}. \\
\\
WASP-39 is a $V=12.11$ star located in an equatorial region of sky
monitored by both WASP instruments, significantly increasing the
observing coverage on the target. In January 2009, the SuperWASP-N
telescope underwent a system upgrade that improved our control over
the main sources of red noise, such as temperature-dependent focus
changes \citep{Barros11}. This upgrade yielded data of unprecedented
high quality, and increased the number of planet candidates, flagged
in the archive, in particular those with longer period and lower mass
(e.g. WASP-38b, \citealt{Barros11}; and WASP-39b).

WASP-39 was routinely observed between 2006 July 1 to 2010 July 26,
with a total of 11 WASP fields and 40531 photometric points. Over the
five WASP seasons only three points were taken in 2006 and none during
2007. However, the same field was observed again in 2008, 2009 and
2010 after both WASP telescopes began observing an overlapping
equatorial region.

\begin{figure}
   \centering
   \includegraphics[width=0.5\textwidth]{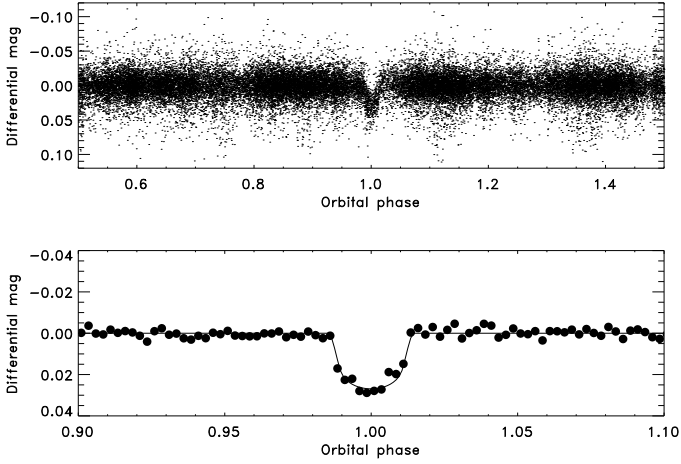}
   \caption{{\em Upper panel}: Discovery light curve of WASP-39b phase folded
    on the ephemeris given in Table~\ref{sys-params3}.
    {\em Lower panel}: binned WASP-39b light curve. Black-solid line, is the best-fit
     transit model estimated using the formalism from
     \citet{Mandel02}.}
              \label{waspLC}
    \end{figure}

All data were processed with a custom-built reduction pipeline
described in \citet{Pollacco06}. The resulting light curves were
analysed using our implementation of the Box Least-Squares and SysRem
detrending algorithms (see \citealt{Cameron06}; \citealt{Kovacs02};
\citealt{Tamuz05}), to search for signatures of planetary transits.
When combined, the SuperWASP light curves showed a characteristic
periodic dip in brightness with a period of $P=4.055$\,days, duration
$T_{14}\sim168$\,minutes, and a depth $\sim25$\,mmag.
Figure~\ref{waspLC} shows the discovery photometry of WASP-39b phase
folded on the above period, along with the binned phased light curve.
A total of 10 partial or full transits were observed, with an
improvement in $\chi^2$ of the box-shaped model over the flat light
curve of $\Delta\chi^2=825$. We evaluated a signal-to-(red) noise
value of the data following \citet{Pont06}, and determined
$SN_{red}=11.65$.

\begin{table}
\caption[]{Photometric properties of WASP-39. The broad band magnitudes and proper motion are 
obtained from the NOMAD~1.0 catalogue.}
\label{table1}
\begin{center}
\begin{tabular}{cc}
\hline
\hline
 Parameter    & WASP-39 \\
 \hline\\
${\rm RA (J2000)}$     & 14:29:18.42        \\ 
${\rm Dec (J2000)}$   & $-$03:26:40.1         \\
                                  &             \\
${\rm B}$                   &  12.93$\pm$0.25 \\
${\rm V}$                   &  12.11$\pm $0.13   \\
${\rm I}$                    &  11.34$\pm $0.08   \\
${\rm J}$                    &  10.663$\pm $0.024   \\
${\rm H}$                  &  10.307$\pm $0.023   \\
${\rm K}$                  &  10.202$\pm $0.023   \\
\mura (mas/yr)         &   $-$12.2 $\pm$3.3           \\
\mudec   (mas/yr)    &   2.8$\pm$3.5           \\
 &             \\
\hline\\
\end{tabular}
\end{center}
\end{table}

\begin{figure}
   \centering
   \includegraphics[width=0.5\textwidth]{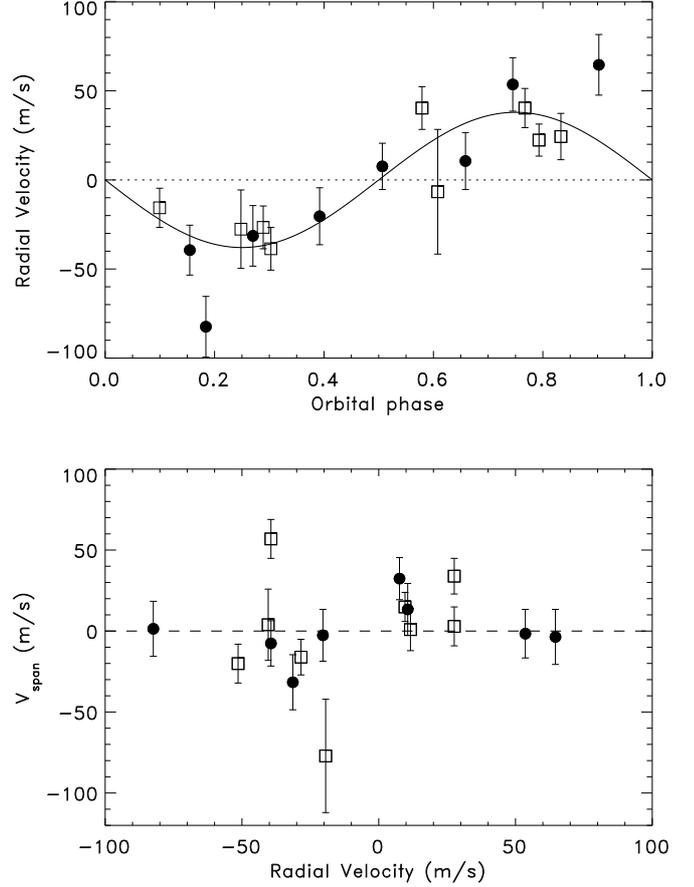}
   \caption{{\em Upper panel}: Phase folded radial velocity
     measurements of WASP-39 obtained combining data from SOPHIE
     (filled-circles) and CORALIE (open-squares) spectrographs.
     Superimposed is the best-fit model RV curve with parameters from
     Table~\ref{sys-params3}. The centre-of-mass velocity for each
     data set was subtracted from the RVs ($\gamma_{\rm SOPHIE} =
     -58.4826$\,\kms and $\gamma_{\rm CORALIE} = -58.4708$\,\kms). {\em
       Lower panel}: we show the bisector span measurements as a
     function of radial velocity, values are shifted to a zero-mean
     ($<V_{span}>_{SOPHIE} = -0.032$\,\kms, $<V_{span}>_{CORALIE}= -0.051$\,\kms).
     The bisector span shows no significant
    variation nor correlation with the RVs, suggesting that the signal
    is mainly due to Doppler shifts of the stellar lines rather than
    stellar profile variations due to stellar activity or a blended
    eclipsing binary.}
    \label{RVfup}%
    \end{figure}

\subsection{Spectroscopic follow-up}

WASP-39 was first observed during our follow-up campaign in April 2010
at Observatoire de Haute-Provence (OHP). During our program we have
obtained follow-up spectroscopy and established the planetary nature
of WASP-39b together with three additional systems: WASP-37b
\citep{Simpson11}, WASP-38b \citep{Barros11} and WASP-40b
\citep{Anderson11a}. Between 2010 April 8 and June 11, we obtained
eight radial velocity measurements for WASP-39 using SOPHIE, the
fiber-fed echelle spectrograph mounted on the 1.93-m telescope at the
OHP (\citealt{Perruchot08}; \citealt{Bouchy09}). We used SOPHIE in
high efficiency mode (R = 40,000) and obtained observations with very
similar signal-to-noise ratio ($\sim$\,30), to minimise systematic
errors arising from the known Charge Transfer Inefficiency effect of
the CCD (\citealt{Bouchy09}), although this is now corrected by the
data reduction software. Wavelength calibration with a Thorium-Argon
lamp were performed every $\sim$\,2 hours, allowing the interpolation
of the spectral drift of SOPHIE ($< 3$\,m s$^{-1}$ per hour; see
\citealt{Boisse2010}). Two 3$\arcsec$~diameter optical fibres were
used, the first centred on the target and the second on the sky to
simultaneously measure the background to remove contamination from
scattered moonlight. During our observations the contribution from
scattered moonlight was negligible as it was well shifted from the
target's radial velocity. Nine additional radial velocity measurements
were obtained using the CORALIE spectrograph mounted on the 1.2-m
Euler Swiss telescope at La Silla, Chile (\citealt{Baranne96};
\citealt{Queloz00}; \citealt{Pepe02}). Observations were obtained with
a signal-to-noise of $\sim$30, during grey/dark time to minimise
moonlight contamination. The data were processed with the SOPHIE and
CORALIE standard data reduction pipelines, respectively. Radial
velocity uncertainties were evaluated, including known systematics
such as guiding and centring errors \citep{Boisse2010}, and wavelength
calibration uncertainties. All spectra were single-lined.

\begin{table} 
\caption{Radial velocity (RV) and line bisector span ($V_{\rm span}$ ) measurements of WASP-39.} 
\label{rv-data} 
\begin{tabular}{cccrl} 
\hline \hline\\
BJD & RV & $\sigma_{\rm RV}$ & V$_{\rm span}$ & Instrument\\ 
$-$2,400,000 & (km s$^{-1}$) & (km s$^{-1}$) & (km s$^{-1}$) &\\ 
\hline\\
55321.7484 & -58.455 & 0.011 & -0.017 & CORALIE \\
55327.7544 & -58.523 & 0.022 & -0.047 & CORALIE\\
55359.5932 & -58.511 & 0.011 & -0.067 & CORALIE\\
55361.6552 & -58.502 & 0.035 & -0.128 & CORALIE\\
55362.5676 & -58.471 & 0.013 & -0.050 & CORALIE\\
55365.5935 & -58.455 & 0.012 & -0.048 & CORALIE\\
55376.6394 & -58.534 & 0.012 & -0.071 & CORALIE\\
55378.6265 & -58.473 & 0.009 & -0.036 & CORALIE\\
55380.6376 & -58.522 & 0.012 &   0.006 & CORALIE\\
\\
\hline
\\
55304.4704 & -58.475 &   0.013 &   0.000 & SOPHIE\\
55305.4374 & -58.429 &   0.015 & -0.034 & SOPHIE\\
55323.4389 & -58.565 &   0.017 & -0.031 & SOPHIE\\
55329.4184 & -58.472 &   0.016 & -0.019 & SOPHIE\\
55331.4301 & -58.522 &   0.014 & -0.040 & SOPHIE\\
55334.4624 & -58.418 &   0.017 & -0.036 & SOPHIE\\
55336.4477 & -58.503 &   0.016 & -0.035 & SOPHIE\\
55368.3955 & -58.514 &   0.017 & -0.064 & SOPHIE\\     
\\
\hline     
\end{tabular}
\end{table}

\begin{figure}
   \centering
   \includegraphics[width=0.5\textwidth]{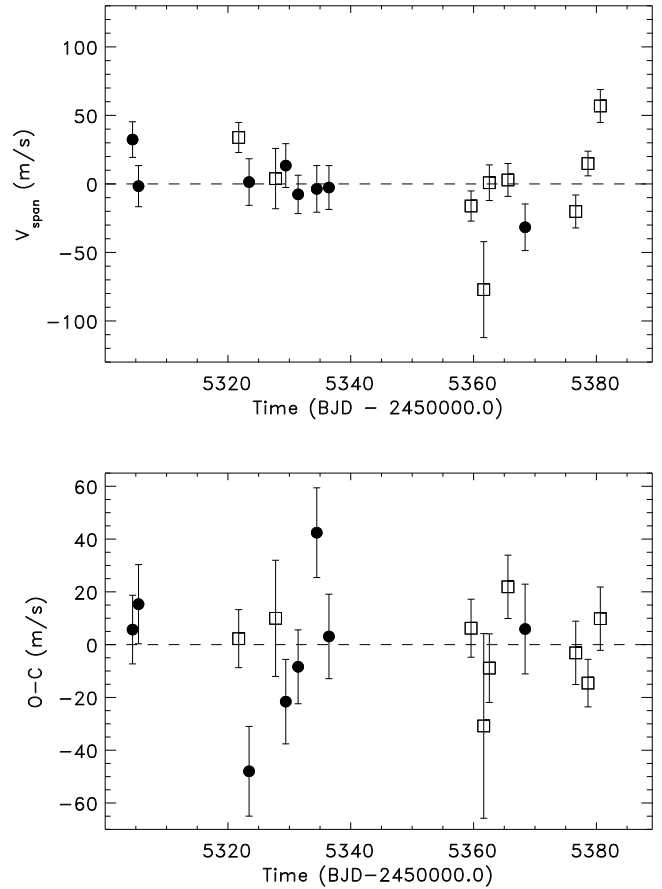}
   \caption{{\em Upper panel}: The bisector span measurements as a
     function of time (BJD--2450000.0), V$_{span}$ values are shifted
     to a zero-mean as in Figure~\ref{RVfup}. {\em Lower panel}:
     Residuals from the RV orbital fit plotted against time.}
    \label{Vspan}%
    \end{figure}

We computed the radial velocities from a weighted
cross-correlation of each spectrum with a numerical mask of
spectral type G2, as described in \citet{Baranne96} and
\citet{Pepe02}. The cross-correlation with masks of different
spectral types (F0, K5 and M5) produced similar radial velocity
variation, rejecting a blended eclipsing system of stars with
unequal masses as a possible cause of the variation.
Radial velocity measurements and line bisector (V$_{span}$) are given
in Table~\ref{rv-data}, and plotted with the best-fit Keplerian model
in Figure~\ref{RVfup}. SOPHIE data are plotted as filled circles and
CORALIE data as open squares, and both data sets are offset with
respect to the radial velocity zero point, $\gamma_{\rm SOPHIE}$ and
$\gamma_{\rm CORALIE}$, respectively (see Table~\ref{sys-params3}). No
significant correlation is observed between the radial velocity and
the line bisector, suggesting the signal's origin as planetary rather
than due to a blended eclipsing binary system or to stellar activity
(see \citealt{Queloz01}). The $RMS$ for SOPHIE and CORALIE radial
velocity residuals to the best-fit model are $RMS_{\rm SOPHIE} =
26$\,m s$^{-1}$ and $RMS_{\rm CORALIE}= 16$\,m s$^{-1}$, with the
higher $RMS$ value of the SOPHIE measurements dominated by two
discrepant measurements, $RV_1 = 48$\,m s$^{-1}$ and $RV_2=43$\,m
s$^{-1}$. Eliminating the most discrepant of these, the $RMS_{\rm
  SOPHIE}$ becomes 19\,m s$^{-1}$, comparable to that of CORALIE. We
also plot in Figure~\ref{Vspan} the bisector span measurements as a
function of time; and lower panel, the residuals from the RV orbital
fit against time. The bisector span values are shifted to a zero mean
($<V_{span}>_{SOPHIE} = -0.032$\,\kms, $<V_{span}>_{CORALIE}=
-0.051$\,\kms), to better compare the two data sets.

\begin{figure}
   \centering
   \includegraphics[width=0.5\textwidth]{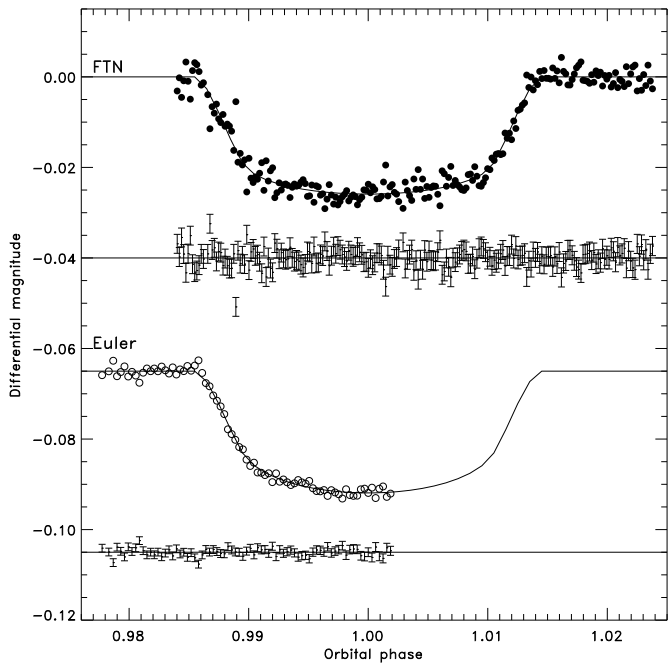}
   \caption{FTN $z-$band and Euler $r-$band follow-up high
     signal-to-noise photometry of WASP-39b during the transit. The
     Euler light curve has been ofsetted from zero by an arbitrary amount for clarity.
     The data are phase-folded on the ephemeris from
     Table~\ref{sys-params3}. Superposed (black-solid line) is the
     best-fit transit model estimated using the formalism from
     \citet{Mandel02}. Residuals from the fit are displayed underneath.}
              \label{photFU}
    \end{figure}

\subsection{Photometric follow-up}
To allow more accurate light curve modelling and thus refine the
photometric parameters, we obtained two high signal-to-noise transit
light curves of WASP-39b. All photometric data presented here are
available from the NStED database
\footnote{http://nsted.ipac.caltech.edu}. The first full transit was
observed with the LCOGT\footnote{http://lcogt.net} 2m Faulkes
Telescope North (FTN) on Haleakala, Maui Hawai'i, on the night of 2010
May 18. The fs03 Spectral Instruments camera was used with a $2 \times
2$ binning mode giving a field-of-view of $10\arcmin \times 10\arcmin$
and a pixel scale of 0.303\arcsec/pixel. Data were taken through a
Pan-STARRS-z filter and the telescope was defocussed during
observations to prevent saturation and allow longer exposure times to
be used. The observations were pre-processed using the WASP Pipeline
\citep{Pollacco06} to perform master bias and flat construction,
debiassing and flatfielding. Aperture photometry was performed with
the the DAOPHOT package \citep{Stetson1987} within the
IRAF\footnote{IRAF is distributed by the National Optical Astronomy
  Observatory, which is operated by the Association of Universities
  for Research in Astronomy, Inc., under cooperative agreement with
  the National Science Foundation.} environment using a 13 pixel
radius aperture. The differential photometry was performed relative to
7 comparison stars that were within the FTN field of view. Additional
high signal-to-noise photometry was obtained in the Gunn\,$r$ filter
with the Euler-Swiss telescope on 2010 July 9, only covering a partial
transit. Conditions were variable, with seeing ranging from 0.6
\arcsec ~to 1.7 \arcsec. The Euler telescope employs an absolute
tracking system, which keeps the star on the same pixel during the
observation by matching the point sources in each image with a
catalogue, and adjusting the telescope pointing between exposures to
compensate for drifts. Observations were obtained with a slightly
defocused (0.1 mm) telescope. All images were corrected for bias and
flat field effects and the light curve was obtained by performing
relative aperture photometry of the target and one bright reference
star.

Both the FTN and Euler light curves are shown in Figure~\ref{photFU}.
The photometry confirms the transit which phases with the ephemeris
derived from the WASP discovery photometry.

\section{Results}

\subsection{Stellar parameters}\label{sec:specsynth}
A total of nine individual CORALIE spectra of WASP-39 were co-added to
produce a single spectrum with a typical S/N ratio of around 50. The
standard pipeline reduction products were used in the analysis. To
improve the line profile fitting for equivalent width measurements,
the spectrum was smoothed using a Gaussian width $\sigma = 0.05$\,\AA.
For the \vsini\ determination the unsmoothed spectrum was used.

Our analysis was performed using the methods given in
\citet{Gillon09}. The \halpha\ line was used to determine the
effective temperature (\teff), while the Na {\sc i} D and Mg {\sc i} b
lines were used as surface gravity (\logg) diagnostics. The
atmospheric parameters obtained from the analysis are listed in
Table~\ref{wasp39-params}. The elemental abundances were determined
from equivalent width measurements of several clean and unblended
lines. A value for microturbulence (\mictrb) was determined from the
Fe~{\sc i} lines using the method of \cite{Magain1984}. Quoted error
estimates include those given by the uncertainties in \teff, \logg\
and \mictrb, as well as the scatter due to measurement and atomic data
uncertainties.

The projected stellar rotation velocity (\vsini) was determined by
fitting the profiles of several unblended Fe~{\sc i} lines. A value
for macroturbulence (\mactrb) of $2.1\pm 0.3$\,\kms\ was assumed,
based on the tabulation by \cite{Gray2008} and an instrumental FWHM of
$0.11\pm 0.01$\,\AA, determined from the telluric lines around
6300~\AA. A best-fitting value of $\vsini = 1.4\pm 0.6$\,\kms\ was
obtained. The stellar mass \mstar~and radius~\rstar~were estimated
using the calibration of \citet{Torres2010}.

\begin{table}
\caption{Stellar parameters of WASP-39 from Spectroscopic Analysis.}
\label{wasp39-params}
\centering
\begin{tabular}{cc} 
\hline\hline
Parameter  & Value \\ 
\hline
\teff      & 5400 $\pm$ 150 K \\
\logg      & 4.4 $\pm$ 0.2 \\
\mictrb    & 0.9 $\pm$ 0.2 \kms \\
\vsini     & 1.4 $\pm$ 0.6 \kms \\
{[Fe/H]}   &$-$0.12 $\pm$ 0.10 \\
{[Na/H]}   &$-$0.04 $\pm$ 0.10 \\
{[Mg/H]}   &   0.06 $\pm$ 0.11 \\
{[Al/H]}   &   0.01 $\pm$ 0.08 \\
{[Si/H]}   &   0.04 $\pm$ 0.08 \\
{[Ca/H]}   &   0.01 $\pm$ 0.14 \\
{[Sc/H]}   &   0.02 $\pm$ 0.19 \\
{[Ti/H]}   &$-$0.03 $\pm$ 0.10 \\
{[V/H]}    &$-$0.08 $\pm$ 0.17 \\
{[Cr/H]}   &$-$0.07 $\pm$ 0.10 \\
{[Mn/H]}   &$-$0.03 $\pm$ 0.20 \\
{[Co/H]}   &$-$0.10 $\pm$ 0.12 \\
{[Ni/H]}   &$-$0.06 $\pm$ 0.09 \\
log A(Li)  &   $<$ 0.9 \\
Mass       &   0.93 $\pm$ 0.09 \msun \\
Radius     &   1.00 $\pm$ 0.25 \rsun \\
Sp. Type   &   G8 \\
Distance   &   230 $\pm$ 80 pc \\ 
\hline
\end{tabular}
\vspace{+0.2cm}
\newline {\bf Note:} Mass and Radius estimate using the
\cite{Torres2010} calibration. Spectral Type estimated from \teff\
using the table in \cite{Gray2008}.
\end{table}

The non-detection of lithium in the spectrum, the low rotation rate
implied by the {\vsini} and lack of stellar activity (shown by the
absence of Ca {\sc ii} H and K emission) all indicate that the star is
relatively old. Unfortunately, the gyrochronological age estimate from
the \citet{Barnes2007} relation ($\sim 5^{+20}_{-4}$\,Gyr) can only
provide a weak constraint on the age of WASP-39.

\begin{figure}
   \centering
   \includegraphics[angle=-90,width=0.5\textwidth]{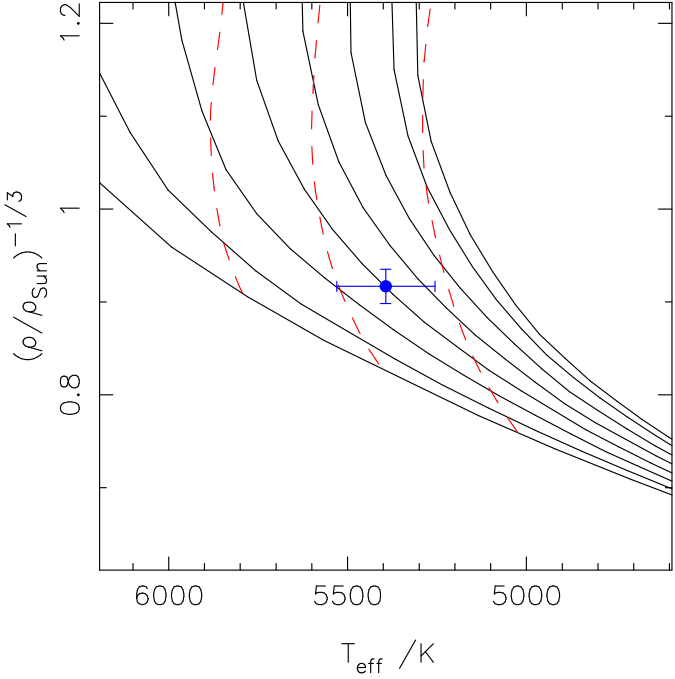}
   \caption{Isochrone tracks from \citet{Demarque04} for WASP-39 
using the best-fit metallicity of \feh~$=- 0.12$\,dex and stellar density 1.297$\,\rho_{\odot}$.
     Solid lines are for isochrones of, left to right: 1.0, 3.0, 6.0, 9.0, 12.0, 15.0, 18.0, and 20.0\,Gyr.
     Dashed lines are for mass tracks of, left to right: 1.0, 0.9, 0.8\,$\msun$.}
              \label{iso}
    \end{figure}

The stellar density of $\rho_{\star} =1.297^{+0.080}_{-0.074}$\,\rhoj~
obtained from the MCMC analysis was used together with the stellar
temperature and metallicity values, derived from spectroscopy, in an
interpolation of the Yonsei-Yale stellar evolution tracks
\citep{Demarque04}, as shown in Figure~\ref{iso}. Using the best-fit
metallicity of \feh~$=- 0.12$, we obtain a mass for WASP-39 of
$0.86\pm0.05\,\msun$ and a stellar age of 9$^{+3}_{-4}$\,Gyr, in
agreement with the gyrochronological age and a more accurate estimate.
The mass obtained from the YY-isochrone fit is somewhat less than the
value from the \citet{Torres2010} calibration (as was also found in
the analysis of WASP-37 stellar parameters; \citealt{Simpson11}), but
their 1--$\sigma$ errors overlap.

\subsection{Planetary parameters}
The planetary properties were determined using a simultaneous
Markov-Chain Monte Carlo (MCMC) analysis including the WASP
photometry, the follow-up FTN and Euler photometry, together with
SOPHIE and CORALIE radial velocity measurements. A detailed
description of the method is given in \citet{Cameron07} and
\citet{Pollacco08}.
The parameters we used in the fit are: the epoch of mid-transit
$T_{0}$, the orbital period $P$, the fractional change of flux
proportional to the ratio of stellar to planet surface areas $\Delta F
= R_{\rm pl}^2/R_{\star}^2$, the transit duration $T_{14}$, the impact
parameter $b$, the radial velocity semi-amplitude $K_{\rm 1}$, the
stellar effective temperature \teff, metallicity \feh, the Lagrangian
elements \secos\ and \sesin~(where $e$ is the eccentricity and
$\omega$ the longitude of periastron), and the systematic offset
velocity $\gamma$. In this particular case we fitted the 2 systematic
velocities $\gamma_{SOPHIE}$ and $\gamma_{CORALIE}$ to allow for
instrumental offsets between the two datasets.

Four different sets of solutions were considered: with and without the
main-sequence mass-radius constraint in the case of circular orbits
and orbits with floating eccentricity. For each solution we have
included a linear trend in the systemic velocity, as a free parameter.
However, we find no significant variation. We used the model of
\citet{Claret00,Claret04} for the limb-darkening in the $r$-band, for
both WASP and Euler photometry, and in the $z$-band for FTN
photometry. Due to the low mass of WASP-39b, the radial velocity data
do not offer convincing evidence for an eccentric orbit. We performed
a Lucy \& Sweeney (\citealt{LucySweeney71}, Eq. 27) F-test, which
indicates that there is a 54\% probability that the improvement in the
fit produced by the best-fitting eccentricity could have arisen by
chance if the orbit were truly circular. Moreover, we find that
imposing the main-sequence constraint has little effect on the MCMC
global solution. Thus, we decided to adopt no main-sequence prior and
circular orbit.

From the above parameters, we calculate the mass $M$, radius $R$,
density $\rho$, and surface gravity $\log g$ of the star (which we
denote with subscript $\star$) and the planet (which we denote with
subscript $\rm pl$), as well as the equilibrium temperature of the
planet assuming it to be a black-body $T_{\rm pl,A=0}$ and that energy
is efficiently redistributed from the planet's day-side to its
night-side. We also calculate the transit ingress/egress times $T_{\rm
  12}$/$T_{\rm 34}$, and the orbital semi-major axis $a$. These
calculated values and their 1-$\sigma$ uncertainties from our MCMC
analysis are presented in Table~\ref{sys-params3}. The corresponding
best-fitting transit light curves are shown in Figures~\ref{waspLC},
Figure~\ref{photFU}, and the best-fitting RV curve in
Figure~\ref{RVfup}.

\begin{table} 
\caption{System parameters of WASP-39} 
\label{sys-params3} 
\begin{tabular}{lc} 
\hline 
Parameter (Unit) & Value \\ 
\hline 
\\
$P$ (d) & 4.055259$ \pm 0.000009$\\
$T_{0}$ (HJD) & 2455342.9688$ \pm 0.0002$\\
$T_{\rm 14}$ (d)~$^a$ & 0.1168$ \pm 0.0008$\\
$T_{\rm 12}=T_{\rm 34}$ (d) \medskip & 0.0179$ \pm 0.0009$\\
$\Delta F=\rpl^{2} / \rstar^{2}$ \medskip & 0.0211$^{+ 0.0003}_{- 0.0004}$\\
$b$ \medskip & 0.441$^{+ 0.036}_{- 0.043}$\\
$i$ ($^\circ$) \medskip & 87.83$^{+ 0.25}_{- 0.22}$\\
$K_{\rm 1}$ (m s$^{-1}$) & 38$ \pm 4$\\
$\gamma_{\rm SOPHIE}$  (\kms) & -58.4826$ \pm 0.0004$\\
$\gamma_{\rm CORALIE}$ (\kms) & -58.4708$ \pm 0.0004$\\
$e$ & 0 (fixed)\\
\mstar~(\msun) & 0.93$ \pm 0.03$\\
\rstar~(\rsun) & 0.895$ \pm 0.023$\\
$\log g_{\star}$ (cgs) \medskip& 4.503$ \pm 0.017$\\
\rhostar~(\rhosun) \medskip & 1.297$^{+ 0.082}_{- 0.074}$\\
\mpl~(\mj) & 0.28$ \pm 0.03$\\
\rpl~(\rj) \medskip& 1.27$ \pm 0.04$\\
$\log g_{\rm pl}$ (cgs) \medskip & 2.610$^{+ 0.047}_{- 0.053}$\\
\rhopl~(\rhoj) & 0.14$ \pm 0.02$\\
$a$ (AU)  \medskip& 0.0486$ \pm 0.0005$\\
$T_{\rm pl, A=0}$ (K) & 1116$^{+ 33}_{- 32}$\\
\\ 
\hline 
\multicolumn{2}{l}{$^{a}$ $T_{\rm 14}$: time between 1$^{st}$ and 4$^{th}$ contact} 
\end{tabular} 
\end{table}

\section{Discussion}
We report the discovery of a new transiting extrasolar planet,
WASP-39b. A simultaneous fit to transit photometry and radial velocity
measurements gives a planetary mass of $0.28 \pm 0.03$\,\mj~ and a
radius of $1.27 \pm 0.040$\,\rj~which yields a planetary density of
$0.141\pm 0.02$\,\rhoj. Thus, WASP-39b is the third least dense planet
identified by a ground-based transit survey. Only WASP-17b
(\citealt{Anderson10a}, $\rho_{W17} = 0.06$\,\rhoj), and WASP-31b
($\rho_{W31} = 0.132$\,\rhoj,~\citealt{Anderson10b}) have lower
densities. However, they are slightly more massive planets
($M_{W17}=0.49$\,\mj~ and $M_{W31}=0.48$\,\mj), but highly irradiated,
with larger and hotter host stars (\citealt{Anderson10a};
\citealt{Anderson10b}). This implies higher planet equilibrium
temperatures for WASP-17b and WASP-31b compared to WASP-39b. We find
WASP-39b to have a highly inflated radius ($\rpl = 1.27$\,\rj), more
than 20\% larger than the \rpl~obtained by comparison with the
\citet{Fortney07} and the \citet{Baraffe08} models for a coreless
planet of a similar mass, orbital distance and stellar age. For
example, tables presented in \citet{Fortney07} predict a maximum
radius of $\sim1.05$\,\rj~ for a 0.24\,\mj~planet orbiting at 0.045 AU
from a 4.5 Gyr solar-type star. In addition, WASP-39 is smaller,
cooler and probably older than the Sun. Thus, a radius of 1.27\,\rj~is
clearly too large for these models. The fact that we do not detect any
eccentricity, and that the age of WASP-39b's host star is
$\sim9$\,Gyr, suggests that it is unlikely that recent tidal
circularisation and dissipation could be a cause of the large radius
of WASP-39b (\citealt{Leconte10}; \citealt{Hansen10}). The low
metallicity (${\rm \feh} = -0.12 \pm 0.1$) of the WASP-39b host star
supports the expected low planetary core-mass. However, this will only
marginally explain the highly inflated radius of WASP-39b. Hence, this
leads to the hypothesis that some additional physics is at play
\citep{Fortney09}. An interior heat source, replacing the radiated
heat from gravitational contraction, is needed for the planet to reach
thermal equilibrium with a larger radius than theoretically expected.
The Ohmic heating mechanism \citep{Batygin11}, based on the
electro-magnetic interaction betweent atmospheric winds and the
planet's magnetic field, is able to explain the anomalous sizes of
close-in transiting planets. WASP-39b has a low equilibrium
temperature ($T_{\rm pl, A=0} =1116$\,K) and thus appears to belong to
the `pL' class of planets from \citet{Fortney08}. We note that, with
the measured planetary parameters (mass, radius, \teq), WASP-39b
agrees with the coreless models in Figure~4 of \citet{Batygin11}.

\begin{figure}
   \centering
   \includegraphics[width=0.5\textwidth]{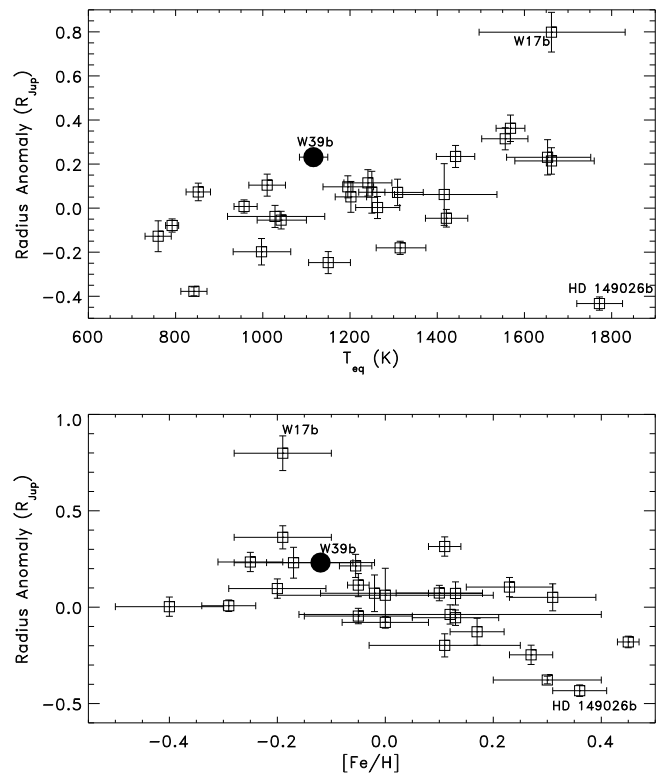}
   \caption{{\em Upper panel}: The radius anomaly $\mathcal R
     =$~$R_{obs}-R_{pred}$, versus equilibrium temperature for the
     known Saturn-mass planets. {\em Lower
       panel}: $\mathcal R$ as function of the stellar metallicity
     \feh\, in dex. WASP-39b is indicated with a filled circle.}
\label{Ranomaly}
\end{figure}

Of the known transiting systems, WASP-39b joins an increasing number
of recently discovered exoplanets with Saturn-like masses. With an
increasingly large sample of well-characterised systems, we can begin
to make statistical inferences as to the physical reasons behind their
diverse nature. \citet{Enoch10} and Anderson et al. (2011 in prep.),
show that the radii of known low-mass (0.15--0.6\,\mj) giant planets
strongly correlate with equilibrium temperature and host-star
metallicity. We have investigated how WASP-39b relates to the rest of
the sample of Saturn-mass planets. We calculated the {\em radius\,
  anomaly}~$\mathcal R$, as in \citet{Laughlin11}, for all Saturn-mass
planets included in the list from \citet{Enoch10}, and added the
latest discoveries (for an updated list see the extrasolar planet
encyclopedia http://exoplanet.eu/). In Figure~\ref{Ranomaly} we plot
$\mathcal R$ against the planetary equilibrium temperature (upper
panel), and as a function of the host star metallicity (lower panel).
WASP-39b is indicated with a filled circle. Figure~\ref{Ranomaly}
shows that $\mathcal R$ correlates with the equilibrium temperature of
the planet, as also found by \citet{Laughlin11}. For transiting gas
giant planets the dominant source of energy is a function of the
orbital separation and the spectral type of the host star, and depends
negligibly on the planetary mass and age (with the exception of very
young planets). Thus, the observed $\mathcal R$\,--\,$T_{\rm eq}$
trend supports the relation between the inflation mechanism and the
stellar irradiation flux \citep{Laughlin11}. The masses of planets in
the Saturn-mass range also appear to correlate with their host star
metallicity (\citealt{Guillot06}; \citealt{Burrows07}), such that
low-density planets are found to orbit sub-solar metallicity stars
(for example WASP-21 \feh\,$= -0.4$, \citealt{Bouchy10}), while higher
density planets orbit stars with super-solar metallicity (for example
WASP-29 \feh\,= 0.11 dex, \citealt{Hellier10}, HD 149026 \feh\,= 0.36 dex,
\citealt{Sato05}, and Kepler-9 \feh\,= 0.12 dex, \citealt{Holman10}).
Figure~\ref{Ranomaly} shows the observed trend between $\mathcal
R$\,--\,\feh,~and highlights the large diversity of planetary internal
structure and composition. WASP-39's metallicity (\feh\,$= -0.12$\,dex)
strengthens this correlation possibly supporting the core-accretion
scenario for planet formation (\citealt{Guillot06};
\citealt{Hartman09}; \citealt{Bouchy10}). The Ohmic dissipation
mechanism \citep{Batygin11}, coupled to the diversity of heavy element
abundances (planetary structural evolution), could provide a universal
mechanism able to slow down planetary cooling and thus the shrinking
of the radius, possibly explaining the observed correlation between
planetary radii and equilibrium temperature. Further progress may be
achieved with a better understanding of the structure and atmospheric
composition of transiting planets (clouds, chemistry, etc).

\begin{acknowledgements} 

  The SuperWASP Consortium consists of astronomers primarily from
  Queens University Belfast, St Andrews, Keele, Leicester, The Open
  University, Isaac Newton Group La Palma and Instituto de Astrofsica
  de Canarias. The SuperWASP-N camera is hosted by the Issac Newton
  Group on La Palma and WASPSouth is hosted by SAAO. We are grateful
  for their support and assistance. Funding for WASP comes from
  consortium universities and from the UK's Science and Technology
  Facilities Council. FPK is grateful to AWE Aldermaston for the award
  of a William Penney Fellowship. Based on observations made with
  SOPHIE spectrograph mounted on the 1.9-m telescope at
  Observatoire de Haute-Provence (CNRS), France and at the ESO La
  Silla Observatory (Chile) with the CORALIE Echelle spectrograph
  mounted on the Swiss telescope. The research leading to these
  results has received funding from the European Community's Seventh
  Framework Programme (FP7/2007-2013) under grant agreement number
  RG226604 (OPTICON). FF is grateful to Yilen G\'omez Maqueo Chew for
  proofreading this paper. FF is grateful to the anonymous referee for
  useful comments improving the paper.

 \end{acknowledgements}

\bibliographystyle{aa}
\bibliography{Faedi}

\begin{thebibliography}{68}
\expandafter\ifx\csname natexlab\endcsname\relax\def\natexlab#1{#1}\fi

\bibitem[{{Anderson} {et~al.}(2011{\natexlab{a}}){Anderson}, {Barros},
  {Boisse}, {Bouchy}, {Collier-Cameron}, {Faedi}, {Hebrard}, {Hellier},
  {Lendl}, {Moutou}, {Pollacco}, {Santerne}, {Smalley}, {Smith}, {Todd},
  {Triaud}, {West}, {Wheatley}, {Bento}, {Enoch}, {Gillon}, {Maxted},
  {McCormac}, {Queloz}, {Simpson}, \& {Skillen}}]{Anderson11a}
{Anderson}, D.~R., {Barros}, S.~C.~C., {Boisse}, I., {et~al.}
  2011{\natexlab{a}}, ArXiv:1101.4643

\bibitem[{{Anderson} {et~al.}(2010{\natexlab{a}}){Anderson}, {Collier Cameron},
  {Hellier}, {Lendl}, {Lister}, {Maxted}, {Queloz}, {Smalley}, {Smith},
  {Triaud}, {West}, {Brown}, {Gillon}, {Pepe}, {Pollacco}, {Segransan},
  {Street}, \& {Udry}}]{Anderson10b}
{Anderson}, D.~R., {Collier Cameron}, A., {Hellier}, C., {et~al.}
  2010{\natexlab{a}}, arXiv:1011.5882

\bibitem[{{Anderson} {et~al.}(2010{\natexlab{b}}){Anderson}, {Hellier},
  {Gillon}, {Triaud}, {Smalley}, {Hebb}, {Collier Cameron}, {Maxted}, {Queloz},
  {West}, {Bentley}, {Enoch}, {Horne}, {Lister}, {Mayor}, {Parley}, {Pepe},
  {Pollacco}, {S{\'e}gransan}, {Udry}, \& {Wilson}}]{Anderson10a}
{Anderson}, D.~R., {Hellier}, C., {Gillon}, M., {et~al.} 2010{\natexlab{b}},
  ApJ, 709, 159

\bibitem[{{Anderson} {et~al.}(2011{\natexlab{b}}){Anderson}, {Smith},
  {Lanotte}, {Barman}, {Campo}, {Collier Cameron}, {Gillon}, {Harrington},
  {Hellier}, {Maxted}, {Queloz}, {Triaud}, \& {Wheatley}}]{Anderson11b}
{Anderson}, D.~R., {Smith}, A.~M.~S., {Lanotte}, A.~A., {et~al.}
  2011{\natexlab{b}}, arXiv:1101.5620

\bibitem[{{Baraffe} {et~al.}(2008){Baraffe}, {Chabrier}, \&
  {Barman}}]{Baraffe08}
{Baraffe}, I., {Chabrier}, G., \& {Barman}, T. 2008, A\&A, 482, 315

\bibitem[{{Baraffe} {et~al.}(2010){Baraffe}, {Chabrier}, \&
  {Barman}}]{Baraffe10}
{Baraffe}, I., {Chabrier}, G., \& {Barman}, T. 2010, Reports on Progress in
  Physics, 73, 016901

\bibitem[{{Baranne} {et~al.}(1996){Baranne}, {Queloz}, {Mayor}, {Adrianzyk},
  {Knispel}, {Kohler}, {Lacroix}, {Meunier}, {Rimbaud}, \& {Vin}}]{Baranne96}
{Baranne}, A., {Queloz}, D., {Mayor}, M., {et~al.} 1996, A\&AS, 119, 373

\bibitem[{{Barnes}(2007)}]{Barnes2007}
{Barnes}, S.~A. 2007, ApJ, 669, 1167

\bibitem[{{Barros} {et~al.}(2011){Barros}, {Faedi}, {Collier Cameron},
  {Lister}, {McCormac}, {Pollacco}, {Simpson}, {Smalley}, {Street}, {Todd},
  {Triaud}, {Boisse}, {Bouchy}, {H{\'e}brard}, {Moutou}, {Pepe}, {Queloz},
  {Santerne}, {Segransan}, {Udry}, {Bento}, {Butters}, {Enoch}, {Haswell},
  {Hellier}, {Keenan}, {Miller}, {Moulds}, {Norton}, {Parley}, {Skillen},
  {Watson}, {West}, \& {Wheatley}}]{Barros11}
{Barros}, S.~C.~C., {Faedi}, F., {Collier Cameron}, A., {et~al.} 2011, A\&A,
  525, A54+

\bibitem[{{Batalha} {et~al.}(2011){Batalha}, {Borucki}, {Bryson}, {Buchhave},
  {Caldwell}, {et~al.}}]{Batalha11}
{Batalha}, N.~M., {Borucki}, W.~J., {Bryson}, S.~T., {et~al.} 2011, ApJ, 729,
  27

\bibitem[{{Batygin} {et~al.}(2011){Batygin}, {Stevenson}, \&
  {Bodenheimer}}]{Batygin11}
{Batygin}, K., {Stevenson}, D.~J., \& {Bodenheimer}, P.~H. 2011, ArXiv e-prints

\bibitem[{{Bodenheimer} {et~al.}(2003){Bodenheimer}, {Laughlin}, \&
  {Lin}}]{Bodenheimer03}
{Bodenheimer}, P., {Laughlin}, G., \& {Lin}, D.~N.~C. 2003, ApJ, 592, 555

\bibitem[{{Bodenheimer} {et~al.}(2001){Bodenheimer}, {Lin}, \&
  {Mardling}}]{Bodenheimer01}
{Bodenheimer}, P., {Lin}, D.~N.~C., \& {Mardling}, R.~A. 2001, ApJ, 548, 466

\bibitem[{{Boisse} {et~al.}(2010){Boisse}, {Eggenberger}, {Santos}, {Lovis},
  {Bouchy}, {H{\'e}brard}, {Arnold}, {Bonfils}, {Delfosse}, {Desort},
  {D{\'{\i}}az}, {Ehrenreich}, {Forveille}, {Gallenne}, {Lagrange}, {Moutou},
  {Udry}, {Pepe}, {Perrier}, {Perruchot}, {Pont}, {Queloz}, {Santerne},
  {S{\'e}gransan}, \& {Vidal-Madjar}}]{Boisse2010}
{Boisse}, I., {Eggenberger}, A., {Santos}, N.~C., {et~al.} 2010, A\&A, 523,
  A88+

\bibitem[{{Bouchy} {et~al.}(2010){Bouchy}, {Hebb}, {Skillen}, {Collier
  Cameron}, {Smalley}, {Udry}, {Anderson}, {Boisse}, {Enoch}, {Haswell},
  {H{\'e}brard}, {Hellier}, {Joshi}, {Kane}, {Maxted}, {Mayor}, {Moutou},
  {Pepe}, {Pollacco}, {Queloz}, {S{\'e}gransan}, {Simpson}, {Smith},
  {Stempels}, {Street}, {Triaud}, {West}, \& {Wheatley}}]{Bouchy10}
{Bouchy}, F., {Hebb}, L., {Skillen}, I., {et~al.} 2010, A\&A, 519, A98+

\bibitem[{{Bouchy} {et~al.}(2009){Bouchy}, {H{\'e}brard}, {Udry}, {Delfosse},
  {Boisse}, {Desort}, {Bonfils}, {Eggenberger}, {Ehrenreich}, {Forveille},
  {Lagrange}, {Le Coroller}, {Lovis}, {Moutou}, {Pepe}, {Perrier}, {Pont},
  {Queloz}, {Santos}, {S{\'e}gransan}, \& {Vidal-Madjar}}]{Bouchy09}
{Bouchy}, F., {H{\'e}brard}, G., {Udry}, S., {et~al.} 2009, A\&A, 505, 853

\bibitem[{{Burrows} {et~al.}(2007){Burrows}, {Hubeny}, {Budaj}, \&
  {Hubbard}}]{Burrows07}
{Burrows}, A., {Hubeny}, I., {Budaj}, J., \& {Hubbard}, W.~B. 2007, ApJ, 661,
  502

\bibitem[{{Chabrier} \& {Baraffe}(2007)}]{Chabrier07}
{Chabrier}, G. \& {Baraffe}, I. 2007, ApJ, 661, L81

\bibitem[{{Claret}(2000)}]{Claret00}
{Claret}, A. 2000, A\&A, 363, 1081

\bibitem[{{Claret}(2004)}]{Claret04}
{Claret}, A. 2004, A\&A, 428, 1001

\bibitem[{{Collier Cameron} {et~al.}(2006){Collier Cameron}, {Pollacco},
  {Street}, {Lister}, {West}, {Wilson}, {Pont}, {Christian}, {Clarkson},
  {Enoch}, {Evans}, {Fitzsimmons}, {Haswell}, {Hellier}, {Hodgkin}, {Horne},
  {Irwin}, {Kane}, {Keenan}, {Norton}, {Parley}, {Osborne}, {Ryans}, {Skillen},
  \& {Wheatley}}]{Cameron06}
{Collier Cameron}, A., {Pollacco}, D., {Street}, R.~A., {et~al.} 2006, MNRAS,
  373, 799

\bibitem[{{Collier Cameron} {et~al.}(2007){Collier Cameron}, {Wilson}, {West},
  {Hebb}, {Wang}, {Aigrain}, {Bouchy}, {Christian}, {Clarkson}, {Enoch},
  {Esposito}, {Guenther}, {Haswell}, {H{\'e}brard}, {Hellier}, {Horne},
  {Irwin}, {Kane}, {Loeillet}, {Lister}, {Maxted}, {Mayor}, {Moutou}, {Parley},
  {Pollacco}, {Pont}, {Queloz}, {Ryans}, {Skillen}, {Street}, {Udry}, \&
  {Wheatley}}]{Cameron07}
{Collier Cameron}, A., {Wilson}, D.~M., {West}, R.~G., {et~al.} 2007, MNRAS,
  380, 1230

\bibitem[{{Demarque} {et~al.}(2004){Demarque}, {Woo}, {Kim}, \&
  {Yi}}]{Demarque04}
{Demarque}, P., {Woo}, J., {Kim}, Y., \& {Yi}, S.~K. 2004, ApJ, 155, 667

\bibitem[{{Enoch} {et~al.}(2011){Enoch}, {Cameron}, {Anderson}, {Lister},
  {et~al.}}]{Enoch10}
{Enoch}, B., {Cameron}, A.~C., {Anderson}, D.~R., {Lister}, T.~A., {et~al.}
  2011, MNRAS, 410, 1631

\bibitem[{{Fortney} {et~al.}(2009){Fortney}, {Baraffe}, \&
  {Militzer}}]{Fortney09}
{Fortney}, J.~J., {Baraffe}, I., \& {Militzer}, B. 2009, ArXiv e-prints

\bibitem[{{Fortney} {et~al.}(2008){Fortney}, {Lodders}, {Marley}, \&
  {Freedman}}]{Fortney08}
{Fortney}, J.~J., {Lodders}, K., {Marley}, M.~S., \& {Freedman}, R.~S. 2008,
  ApJ, 678, 1419

\bibitem[{{Fortney} {et~al.}(2007){Fortney}, {Marley}, \& {Barnes}}]{Fortney07}
{Fortney}, J.~J., {Marley}, M.~S., \& {Barnes}, J.~W. 2007, ApJ, 659, 1661

\bibitem[{{Fortney} \& {Nettelmann}(2010)}]{Fortney10}
{Fortney}, J.~J. \& {Nettelmann}, N. 2010, Space Science Reviews, 152, 423

\bibitem[{{Gillon} {et~al.}(2009){Gillon}, {Smalley}, {Hebb}, {Anderson},
  {Triaud}, {Hellier}, {Maxted}, {Queloz}, \& {Wilson}}]{Gillon09}
{Gillon}, M., {Smalley}, B., {Hebb}, L., {et~al.} 2009, A\&A, 496, 259

\bibitem[{{Gray}(2008)}]{Gray2008}
{Gray}, D.~F. 2008, {The Observation and Analysis of Stellar Photospheres}, ed.
  {Gray, D.~F.}

\bibitem[{{Guillot}(2005)}]{Guillot05}
{Guillot}, T. 2005, Annual Review of Earth and Planetary Sciences, 33, 493

\bibitem[{{Guillot} {et~al.}(2006){Guillot}, {Santos}, {Pont}, {Iro}, {Melo},
  \& {Ribas}}]{Guillot06}
{Guillot}, T., {Santos}, N.~C., {Pont}, F., {et~al.} 2006, A\&A, 453, L21

\bibitem[{{Guillot} \& {Showman}(2002)}]{Guillot02}
{Guillot}, T. \& {Showman}, A.~P. 2002, A\&A, 385, 156

\bibitem[{{Hansen}(2010)}]{Hansen10}
{Hansen}, B.~M.~S. 2010, ApJ, 723, 285

\bibitem[{{Hartman} {et~al.}(2011){Hartman}, {Bakos}, {Sato}, {Torres},
  {et~al.}}]{Hartman10}
{Hartman}, J.~D., {Bakos}, G.~{\'A}., {Sato}, B., {Torres}, G., {et~al.} 2011,
  ApJ, 726, 52

\bibitem[{{Hartman} {et~al.}(2009){Hartman}, {Bakos}, {Torres}, {Kov{\'a}cs},
  {Noyes}, {P{\'a}l}, {Latham}, {Sip{\H o}cz}, {Fischer}, {Johnson}, {Marcy},
  {Butler}, {Howard}, {Esquerdo}, {Sasselov}, {Kov{\'a}cs}, {Stefanik},
  {Fernandez}, {L{\'a}z{\'a}r}, {Papp}, \& {S{\'a}ri}}]{Hartman09}
{Hartman}, J.~D., {Bakos}, G.~{\'A}., {Torres}, G., {et~al.} 2009, ApJ, 706,
  785

\bibitem[{{Hebb} {et~al.}(2009){Hebb}, {Collier-Cameron}, {Loeillet},
  {Pollacco}, {H{\'e}brard}, {Street}, {Bouchy}, {Stempels}, {Moutou},
  {Simpson}, {Udry}, {Joshi}, {West}, {Skillen}, {Wilson}, {McDonald},
  {Gibson}, {Aigrain}, {Anderson}, {Benn}, {Christian}, {Enoch}, {Haswell},
  {Hellier}, {Horne}, {Irwin}, {Lister}, {Maxted}, {Mayor}, {Norton}, {Parley},
  {Pont}, {Queloz}, {Smalley}, \& {Wheatley}}]{Hebb09}
{Hebb}, L., {Collier-Cameron}, A., {Loeillet}, B., {et~al.} 2009, ApJ, 693,
  1920

\bibitem[{{H{\'e}brard} {et~al.}(2008){H{\'e}brard}, {Bouchy}, {Pont},
  {Loeillet}, {Rabus}, {Bonfils}, {Moutou}, {Boisse}, {Delfosse}, {Desort},
  {Eggenberger}, {Ehrenreich}, {Forveille}, {Lagrange}, {Lovis}, {Mayor},
  {Pepe}, {Perrier}, {Queloz}, {Santos}, {S{\'e}gransan}, {Udry}, \&
  {Vidal-Madjar}}]{Hebrard08}
{H{\'e}brard}, G., {Bouchy}, F., {Pont}, F., {et~al.} 2008, A\&A, 488, 763

\bibitem[{{Hellier} {et~al.}(2010){Hellier}, {Anderson}, {Collier Cameron},
  {Gillon}, {Lendl}, {Maxted}, {Queloz}, {Smalley}, {Triaud}, {West}, {Brown},
  {Enoch}, {Lister}, {Pepe}, {Pollacco}, {S{\'e}gransan}, \&
  {Udry}}]{Hellier10}
{Hellier}, C., {Anderson}, D.~R., {Collier Cameron}, A., {et~al.} 2010, ApJ,
  723, L60

\bibitem[{{Holman} {et~al.}(2010){Holman}, {Fabrycky}, {Ragozzine}, {Ford},
  {et~al.}}]{Holman10}
{Holman}, M.~J., {Fabrycky}, D.~C., {Ragozzine}, D., {Ford}, E.~B., {et~al.}
  2010, Science, 330, 51

\bibitem[{{Johns-Krull} {et~al.}(2008){Johns-Krull}, {McCullough}, {Burke},
  {Valenti}, {Janes}, {Heasley}, {Prato}, {Bissinger}, {Fleenor}, {Foote},
  {Garcia-Melendo}, {Gary}, {Howell}, {Mallia}, {Masi}, \&
  {Vanmunster}}]{Johns-Krull08}
{Johns-Krull}, C.~M., {McCullough}, P.~R., {Burke}, C.~J., {et~al.} 2008, ApJ,
  677, 657

\bibitem[{{Kov{\'a}cs} {et~al.}(2002){Kov{\'a}cs}, {Zucker}, \&
  {Mazeh}}]{Kovacs02}
{Kov{\'a}cs}, G., {Zucker}, S., \& {Mazeh}, T. 2002, A\&A, 391, 369

\bibitem[{{Latham} {et~al.}(2010){Latham}, {Borucki}, {Koch}, {Brown},
  {Buchhave}, {Basri}, {Batalha}, {Caldwell}, {Cochran}, {Dunham}, {F{\H
  u}r{\'e}sz}, {Gautier}, {Geary}, {Gilliland}, {Howell}, {Jenkins},
  {Lissauer}, {Marcy}, {Monet}, {Rowe}, \& {Sasselov}}]{Latham10}
{Latham}, D.~W., {Borucki}, W.~J., {Koch}, D.~G., {et~al.} 2010, ApJ, 713, L140

\bibitem[{{Laughlin} {et~al.}(2011){Laughlin}, {Crismani}, \&
  {Adams}}]{Laughlin11}
{Laughlin}, G., {Crismani}, M., \& {Adams}, F.~C. 2011, ArXiv e-prints

\bibitem[{{Leconte} {et~al.}(2010){Leconte}, {Chabrier}, {Baraffe}, \&
  {Levrard}}]{Leconte10}
{Leconte}, J., {Chabrier}, G., {Baraffe}, I., \& {Levrard}, B. 2010, A\&A, 516,
  A64+

\bibitem[{{Liu} {et~al.}(2008){Liu}, {Burrows}, \& {Ibgui}}]{Liu08}
{Liu}, X., {Burrows}, A., \& {Ibgui}, L. 2008, ApJ, 687, 1191

\bibitem[{{Lucy} \& {Sweeney}(1971)}]{LucySweeney71}
{Lucy}, L.~B. \& {Sweeney}, M.~A. 1971, AJ, 76, 544

\bibitem[{{Magain}(1984)}]{Magain1984}
{Magain}, P. 1984, A\&A, 134, 189

\bibitem[{{Mandel} \& {Agol}(2002)}]{Mandel02}
{Mandel}, K. \& {Agol}, E. 2002, ApJ, 580, L171

\bibitem[{{Mandushev} {et~al.}(2007){Mandushev}, {O'Donovan}, {Charbonneau},
  {Torres}, {Latham}, {Bakos}, {Dunham}, {Sozzetti}, {Fern{\'a}ndez},
  {Esquerdo}, {Everett}, {Brown}, {Rabus}, {Belmonte}, \&
  {Hillenbrand}}]{Mandushev07}
{Mandushev}, G., {O'Donovan}, F.~T., {Charbonneau}, D., {et~al.} 2007, ApJ,
  667, L195

\bibitem[{{Pepe} {et~al.}(2002){Pepe}, {Mayor}, {Galland}, {Naef}, {Queloz},
  {Santos}, {Udry}, \& {Burnet}}]{Pepe02}
{Pepe}, F., {Mayor}, M., {Galland}, F., {et~al.} 2002, A\&A, 388, 632

\bibitem[{{Perruchot} {et~al.}(2008)}]{Perruchot08}
{Perruchot}, S. {et~al.} 2008, in Society of Photo-Optical Instrumentation
  Engineers (SPIE) Conference Series, Vol. 7014, Society of Photo-Optical
  Instrumentation Engineers (SPIE) Conference Series

\bibitem[{{Pollacco} {et~al.}(2008){Pollacco}, {Skillen}, {Collier Cameron},
  {et~al.}}]{Pollacco08}
{Pollacco}, D., {Skillen}, I., {Collier Cameron}, A., {et~al.} 2008, MNRAS,
  385, 1576

\bibitem[{{Pollacco} {et~al.}(2006){Pollacco}, {Skillen}, {Collier Cameron},
  {Christian}, {Hellier}, {Irwin}, {Lister}, {Street}, {West}, {Anderson},
  {Clarkson}, {Deeg}, {Enoch}, {Evans}, {Fitzsimmons}, {Haswell}, {Hodgkin},
  {Horne}, {Kane}, {Keenan}, {Maxted}, {Norton}, {Osborne}, {Parley}, {Ryans},
  {Smalley}, {Wheatley}, \& {Wilson}}]{Pollacco06}
{Pollacco}, D.~L., {Skillen}, I., {Collier Cameron}, A., {et~al.} 2006, PASP,
  118, 1407

\bibitem[{{Pont} {et~al.}(2006){Pont}, {Zucker}, \& {Queloz}}]{Pont06}
{Pont}, F., {Zucker}, S., \& {Queloz}, D. 2006, MNRAS, 373, 231

\bibitem[{{Queloz} {et~al.}(2000){Queloz}, {Eggenberger}, {Mayor}, {Perrier},
  {Beuzit}, {Naef}, {Sivan}, \& {Udry}}]{Queloz00}
{Queloz}, D., {Eggenberger}, A., {Mayor}, M., {et~al.} 2000, A\&A, 359, L13

\bibitem[{{Queloz} {et~al.}(2001){Queloz}, {Henry}, {Sivan}, {Baliunas},
  {Beuzit}, {Donahue}, {Mayor}, {Naef}, {Perrier}, \& {Udry}}]{Queloz01}
{Queloz}, D., {Henry}, G.~W., {Sivan}, J.~P., {et~al.} 2001, A\&A, 379, 279

\bibitem[{{Sackett}(1999)}]{Sackett99}
{Sackett}, P.~D. 1999, in NATO ASIC Proc. 532: Planets Outside the Solar
  System: Theory and Observations, ed. J.-M. {Mariotti} \& D.~{Alloin}, 189--+

\bibitem[{{Sato} {et~al.}(2005){Sato}, {Fischer}, {Henry}, {Laughlin},
  {Butler}, {Marcy}, {Vogt}, {Bodenheimer}, {Ida}, {Toyota}, {Wolf}, {Valenti},
  {Boyd}, {Johnson}, {Wright}, {Ammons}, {Robinson}, {Strader}, {McCarthy},
  {Tah}, \& {Minniti}}]{Sato05}
{Sato}, B., {Fischer}, D.~A., {Henry}, G.~W., {et~al.} 2005, ApJ, 633, 465

\bibitem[{{Simpson} {et~al.}(2011){Simpson}, {Faedi}, {Barros},
  {et~al.}}]{Simpson11}
{Simpson}, E.~K., {Faedi}, F., {Barros}, S.~C.~C., {et~al.} 2011, AJ, 141, 8

\bibitem[{{Skrutskie} {et~al.}(2006){Skrutskie}, {Cutri}, {Stiening},
  {Weinberg}, {Schneider}, {Carpenter}, {Beichman}, {Capps}, {Chester},
  {Elias}, {Huchra}, {Liebert}, {Lonsdale}, {Monet}, {Price}, {Seitzer},
  {Jarrett}, {Kirkpatrick}, {Gizis}, {Howard}, {Evans}, {Fowler}, {Fullmer},
  {Hurt}, {Light}, {Kopan}, {Marsh}, {McCallon}, {Tam}, {Van Dyk}, \&
  {Wheelock}}]{2MASS}
{Skrutskie}, M.~F., {Cutri}, R.~M., {Stiening}, R., {et~al.} 2006, AJ, 131,
  1163

\bibitem[{{Southworth}(2010)}]{Southworth10}
{Southworth}, J. 2010, MNRAS, 408, 1689

\bibitem[{{Standish}(1995)}]{Standish95}
{Standish}, E.~M. 1995, Highlights of Astronomy, 10, 180

\bibitem[{{Stetson}(1987)}]{Stetson1987}
{Stetson}, P.~B. 1987, PASP, 99, 191

\bibitem[{{Tamuz} {et~al.}(2005){Tamuz}, {Mazeh}, \& {Zucker}}]{Tamuz05}
{Tamuz}, O., {Mazeh}, T., \& {Zucker}, S. 2005, MNRAS, 356, 1466

\bibitem[{{Torres} {et~al.}(2010){Torres}, {Andersen}, \&
  {Gim{\'e}nez}}]{Torres2010}
{Torres}, G., {Andersen}, J., \& {Gim{\'e}nez}, A. 2010, A\&ARv, 18, 67

\bibitem[{{Torres} {et~al.}(2008){Torres}, {Winn}, \& {Holman}}]{Torres08}
{Torres}, G., {Winn}, J.~N., \& {Holman}, M.~J. 2008, ApJ, 677, 1324

\bibitem[{{Zacharias} {et~al.}(2005){Zacharias}, {Monet}, {Levine}, {Urban},
  {Gaume}, \& {Wycoff}}]{Nomadcat}
{Zacharias}, N., {Monet}, D.~G., {Levine}, S.~E., {et~al.} 2005, VizieR Online
  Data Catalog, 1297, 0

\end{thebibliography}

\end{document}